\documentclass[letter]{aa}
\usepackage{graphicx}
\usepackage{txfonts}
\usepackage{hyperref}
\hypersetup{
    colorlinks=true,
    linkcolor=blue,
   citecolor=blue, 
    filecolor=magenta,
   urlcolor=cyan,
}
\begin{document}

\title{Circular polarization of gravitational waves from magnetorotational supernovae}

\author{Shota~Shibagaki\inst{1,2},
Tomoya~Takiwaki\inst{3},
Kei~Kotake\inst{4},
Takami~Kuroda\inst{5},
\and
Tobias~Fischer\inst{2,6}
}

\institute{
Incubator of Scientific Excellence---Centre for Simulations of Superdense Fluids, University of Wrocław, 50-204, Wroc{\l}aw, Poland\\
\email{shota.sbgk@gmail.com}
\and
Institute of Theoretical Physics, Wroc{\l}aw University of Science and Technology, W.~Wyspia${\rm\dot{n}}$skiego 27, 50-370 Wroc{\l}aw, Poland
\and
National Astronomical Observatory of Japan (NAOJ), 2-21-1, Osawa, Mitaka, Tokyo, 181-8588, Japan
\and
Department of Applied Physics, Fukuoka University, 8-19-1, Nanakuma, Jonan, Fukuoka 814-0180, Japan
\and
Max-Planck-Institut f{\"u}r Gravitationsphysik, Am M{\"u}hlenberg 1, D-14476 Potsdam-Golm, Germany
\and
Research Center for Computational Physics and Data Processing, Institute of Physics, Silesian University in Opava, Bezručovo nám. 13, CZ-746-01 Opava, Czech Republic
             }

   \date{\today}

\abstract
{Gravitational waves (GWs) provide a unique probe of the explosion mechanism of 
massive stars
and the evolution of nascent proto-neutron stars (PNSs). 
Magnetorotational explosions are one of the promising noncanonical core-collapse supernova scenarios, and they might be linked to magnetar formation and energetic supernova explosions. However, the GW signatures of such events currently remain incompletely understood.}
{
We investigate the origin and nature of GW polarization arising from a magnetorotational core-collapse model and examine its potential detectability by current GW observatories.
}
{We performed a three-dimensional simulation of general-relativistic magnetohydrodynamics of a rapidly rotating, strongly magnetized $20\,M_\odot$ progenitor, including multi-energy neutrino transport. 
The GW signals were extracted using the standard quadrupole formalism, and their polarization states were analyzed with Stokes parameters.}
{Strong circular polarization emerges along the rotation axis during the early post-bounce phase ($\lesssim 230$ ms after core bounce). The characteristic GW spectrum peaks at $\sim$90~Hz, consistent with the emission at twice the local angular velocity ($\sim$45~Hz) around the PNS 
surface at cylindrical radii of $\sim$50~km. 
These features are attributed to the low-$T/|W|$ instabilities and nonaxisymmetric motions near the PNS and not to the magnetohydrodynamic jets themselves. 
The polarization signals lie within the sensitivity bands of current detectors such as Advanced LIGO, Advanced Virgo, and KAGRA. }
{Our study demonstrates that models in which magnetorotationally driven jets are launched can produce circularly polarized GW signals originating from the inner PNS region. This provides an observational signature that complements previous findings from nonmagnetized rotating models. 
Thus, our novel findings establish that the GW polarization 
is  a promising diagnostic of noncanonical core-collapse supernovae. 
Future third-generation detectors will be crucial to fully exploit this potential.
}
 
    
\authorrunning{S.~Shibagaki et al.}

\maketitle

\section{Introduction}
The gravitational wave (GW) emission from the next Galactic core-collapse supernova (CCSN) is regarded as one of the most promising targets for multimessenger astronomy \citep{vicky21,Szczepaczyk22,Szczepaczyk24}. 
The direct detection of GWs by the LIGO--Virgo--KAGRA collaboration has already marked the beginning of a new era of astronomy, providing unique insights into the mergers of compact binaries such as black holes and neutron stars \citep{GWTC1,GWTC2,GWTC3}. While binary mergers remain the most firmly established GW sources, the detection of GWs from CCSNe would open a direct observational window into the dynamics of the explosion engine, which cannot be accessed by electromagnetic or neutrino observations alone~\citep[see e.g.,][]{Kotake12_ptep,Janka16review,Bmueller16review,Radice18review,Burrows21review,janka25,anders26}.

For the majority of CCSNe, whose canonical explosion energies are of the order of $10^{51}$\,erg, the neutrino-heating mechanism is widely considered as the dominant driver \citep[e.g., ][for collective references therein]{tony20,bernhard25,yama24,Nakamura25}. Numerous multidimensional simulations have predicted GW emission from neutrino-driven models, typically arising from convection, standing accretion shock instability (SASI), and proto-neutron star (PNS) oscillations 
\citep[e.g.,][]{haakon17,kotake_kuroda16,Powell19,Abdikamalov22,Mezzacappa23,michael23,choi24}. The GW signals in this case are largely stochastic, reflecting turbulent flows, and they are thus difficult to predict a priori. However, recent advances in PNS asteroseismology have revealed that characteristic oscillation modes of the PNS may produce nearly universal spectral features (notably, the ramp-up of the quadrupolar $f$- and $g$-modes), which could serve as robust observational signatures~\citep{Torres-Forne19,Torres-Forne21,sotani21}. The possible detection of such features has been gaining increasing attention as a probe of dense-matter physics and explosion dynamics.  

Observational evidence also points to an intriguing subclass of energetic explosions, the so-called hypernovae, with kinetic energies reaching $\sim$10$^{52}$ erg \citep{Nomoto06,taddia18}. These events are often accompanied by broad-lined Type Ic supernovae and long gamma-ray bursts, suggesting that rapid rotation and strong magnetic fields at the stellar core play a central role. For the inner workings, the magnetorotational mechanism is one of the most promising mechanisms. It relies on the extraction of rotational energy of the PNS via the magnetic fields, which often results in the formation of magnetohydrodynamic (MHD) jets. In the context of rapidly rotating collapse, theoretical studies have long predicted distinct GW signatures: a strong bounce signal, the development of the low-$T/|W|$ instability producing quasi-periodic waveforms, and potentially detectable circular polarization that encodes information on rotation and nonaxisymmetric instabilities including SASI \citep{Rampp98,Ott05,Dimmelmeier08,Takiwaki18,Shibagaki20,Shibagaki21,Kuroda25,sophia26}. Recently, increasingly realistic simulations have begun to approach the hypernova regime, including the three-dimensional (3D) special relativistic MHD models of \citet{Obergaulinger20,Obergaulinger22}, which achieve explosion energies approaching $10^{52}$ erg, and the 2D fully general-relativistic (GR) models of \citet{KurodaT24}.

Despite these advances, the GW emission from MHD-driven jet explosions, in particular, its circular polarization properties, has not been comprehensively explored. The treatment of magnetic fields in this context has remained limited. The studies by \citet{Hayama18} and \citet{Shibagaki21} explored the detectability and properties of circular polarization using purely hydrodynamic models and neglected magnetic effects. \citet{Bugli23} and \citet{Shibagaki24} performed MHD simulations of magnetorotational core collapse, but did not investigate GW polarization. In particular, \citet{Bugli23} found the low-$T/|W|$-type GW signals only in hydrodynamic models, as angular momentum transport by magnetic stresses significantly suppressed the rotation of the PNS in their MHD cases. 

Joining these ongoing efforts, we investigate the GW emission from a 3D GRMHD jet model of \citet{Shibagaki24} that incorporates multi-energy neutrino transport. Using a rapidly rotating, strongly magnetized $20 M_\odot$ progenitor, we analyze the GW signals with particular focus on their polarization properties and detectability by current detectors. We show that strong circular polarization arises in this jet-producing model, primarily due to the low-$T/|W|$ instabilities and nonaxisymmetric motions around the PNS, and is not directly generated by the jets themselves. We also discuss the implications for probing the explosion mechanism of noncanonical supernovae through GW observations.

In Sect.~\ref{Sec:Results} we present the post-bounce dynamics and the overall characteristics of the magnetorotational CCSN explosion, followed by an analysis of the GW signal and the polarization properties. 
We discuss the physical origin of the circularly polarized components and identify the dominant emission regions associated with the low-$T/|W|$ instability. 
Sect.~\ref{Sec:Conclusion} summarizes our findings and outlines prospects for future studies. 
The numerical methods and analysis procedures are described in Appendix~\ref{Sec:NumericalSetup}.

\section{Results}
\label{Sec:Results}
First, we briefly revisit the dynamical evolution of our magnetorotational CCSN explosion model. 
After core bounce, the strong differential rotation of the PNS strengthens the magnetic fields by quickly winding up the magnetic field lines along the rotational axis, which results in the MHD jet launch at $t_{\rm{pb}}\sim 60\,{\rm ms}$ (with $t_{\rm pb}$ being the post-bounce time).
The equatorial shock surface reaches 300\,km a few tens of milliseconds after this time and continues to expand. 
At the final simulation time of $t_{\rm{pb}}=545$\,ms, the maximum shock radius reaches $\sim$11000\,km.
Figure 1 illustrates the jet morphology at $t_{\rm pb}\sim100\,{\rm ms}$.
It also shows the normalized density deviation on the equatorial plane, $(\rho-\langle\rho\rangle)/\langle\rho\rangle$, where $\langle \rangle$ indicates the angular average with respect to the azimuthal angle.
The large-scale deviation in density is clearly visible, which is most likely to be attributed to the growth of the low-$T/|W|$ instability \citep{Ott05} or spiral SASI \citep{Blondin03,remi17,Walk23}. 
We discuss this point further below. 

The central PNS is strongly magnetized, with the PNS surface magnetic field strength reaching several times $10^{14}$~Gauss, and high rotation rates of several 100~rad~s$^{-1}$~\citep[see Fig.~3 in ][]{Shibagaki24}. Consequently, the associated Maxwell stress decelerates the PNS rotation.
In our model, the average angular velocity of the PNS starts to decrease around $t_{\rm{pb}}\sim 220$~ms, and it vanishes at around $t_{\rm{pb}}\sim 350$\,ms.

\begin{figure}
\centering
\includegraphics[width=0.4\textwidth]{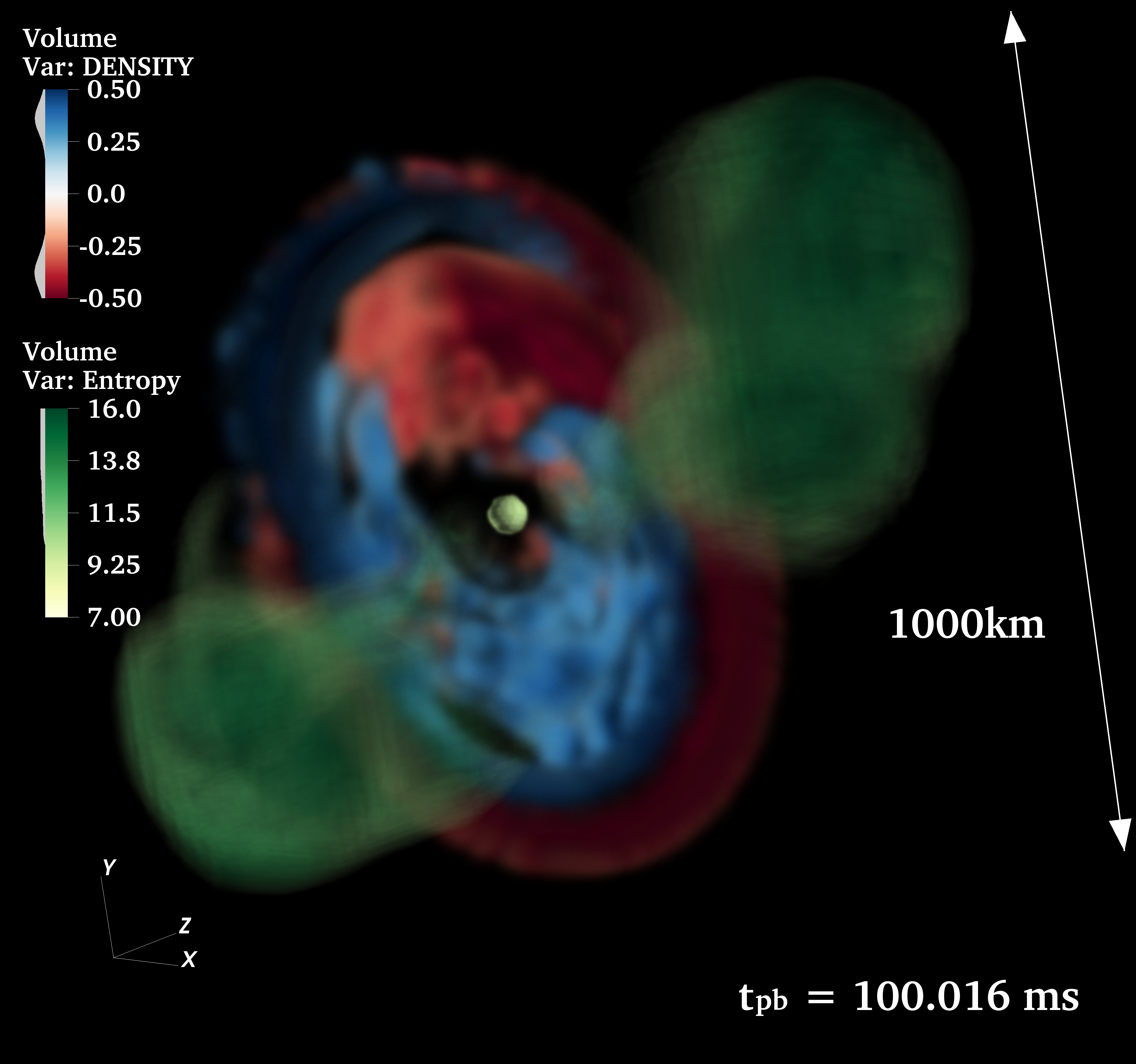}
\caption{3D entropy plot for MHD jets (green region) at $t_{\mathrm{pb}}=100$\,ms. The red and blue region around the center is the normalized density deviation from the angle-averaged density on the equatorial plane. The central light yellow sphere is the PNS. \label{fig:3Dent}}
\end{figure}

\begin{figure}
\centering
\includegraphics[width=0.44\textwidth]{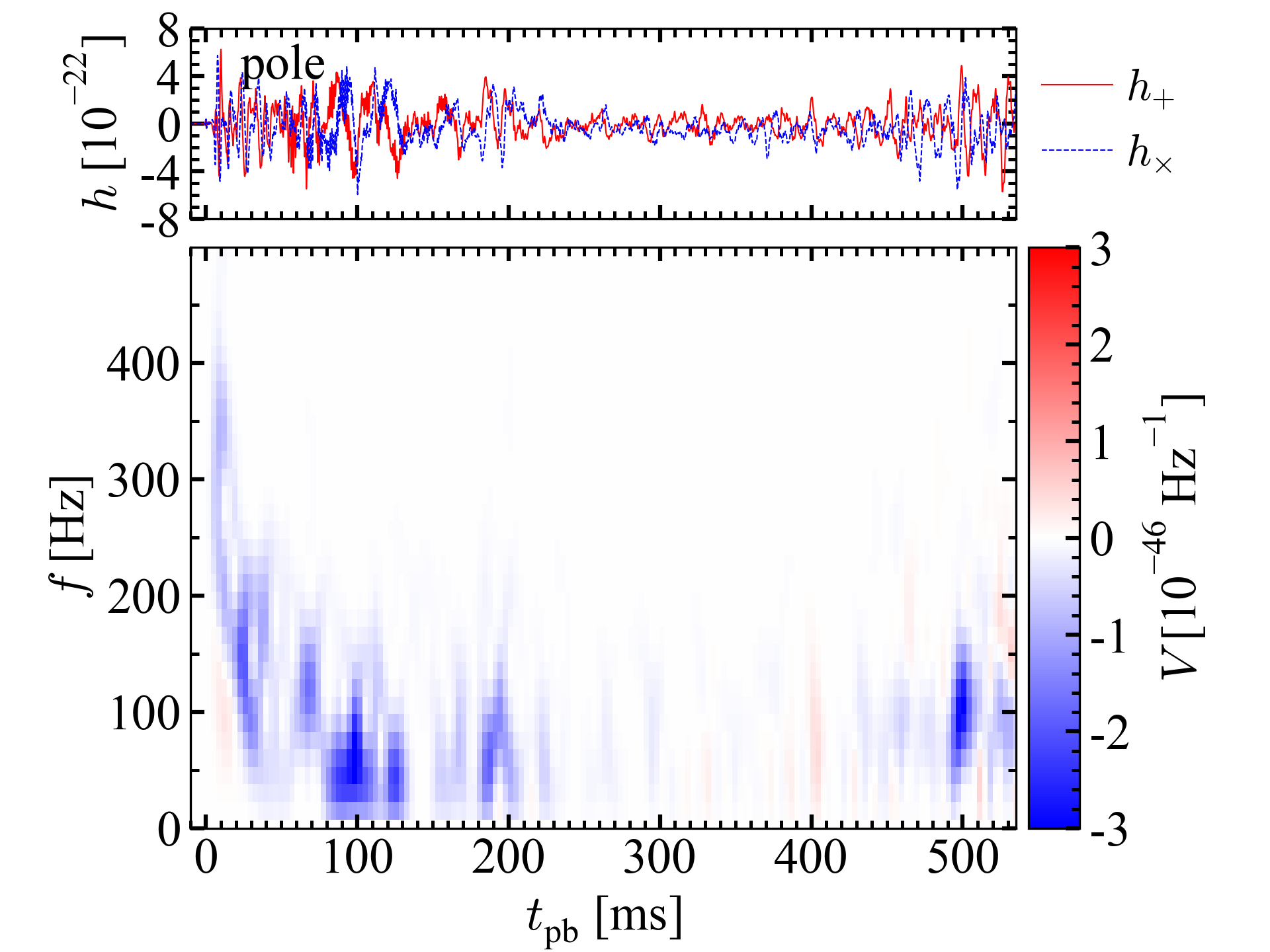}
\caption{GW strains of plus (solid red) and cross (dashed blue) modes (top) and spectrogram of its $V$-mode power spectrum (bottom) seen along the pole at a source distance of 10\,kpc. \label{fig:specV}}
\end{figure}

The upper panel of Fig.~\ref{fig:specV} shows the plus and cross modes of the GW strains, $h_{+}$ and $h_{\times}$, respectively, observed along the initial rotational axis, i.e., the direction of the north pole at a source distance of 10\,kpc.
The lower panel shows its $V$-mode spectrogram.
We find the coherent large $V$-mode amplitude at $t_{\rm{pb}}<230\,{\rm ms}$, which indicates strong circular polarization of the GW.
Following \citet{Bugli23}, we compared this blue region in Fig.~\ref{fig:specV} to twice the maximum rotation frequency.
As was indicated by \citet{Bugli23} for the usual characteristic strain, we confirm here that for $t_{\mathrm{pb}}<230$\,ms, this blue region is always below twice the maximum rotation frequency, indicating that this feature originates from rotation.
The moment-of-inertia–weighted average rotation rate of the PNS decreases rapidly at $220\,{\rm ms}<t_{\rm{pb}}<340\,{\rm ms}$, after which the PNS hardly rotates at all~\citep{Shibagaki24}.
The $V$-mode amplitude is in a quiescent phase during $220\,{\rm ms} < t_{\rm{pb}} < 400\,{\rm ms}$ and irregularly shows both positive and negative values afterward.
This is a clear indication that the $V$-mode amplitude does not stem from the rotational effect after $t_{\rm{pb}}\sim 220$\,ms alone. 

\begin{figure}
\centering
\includegraphics[width=0.44\textwidth]{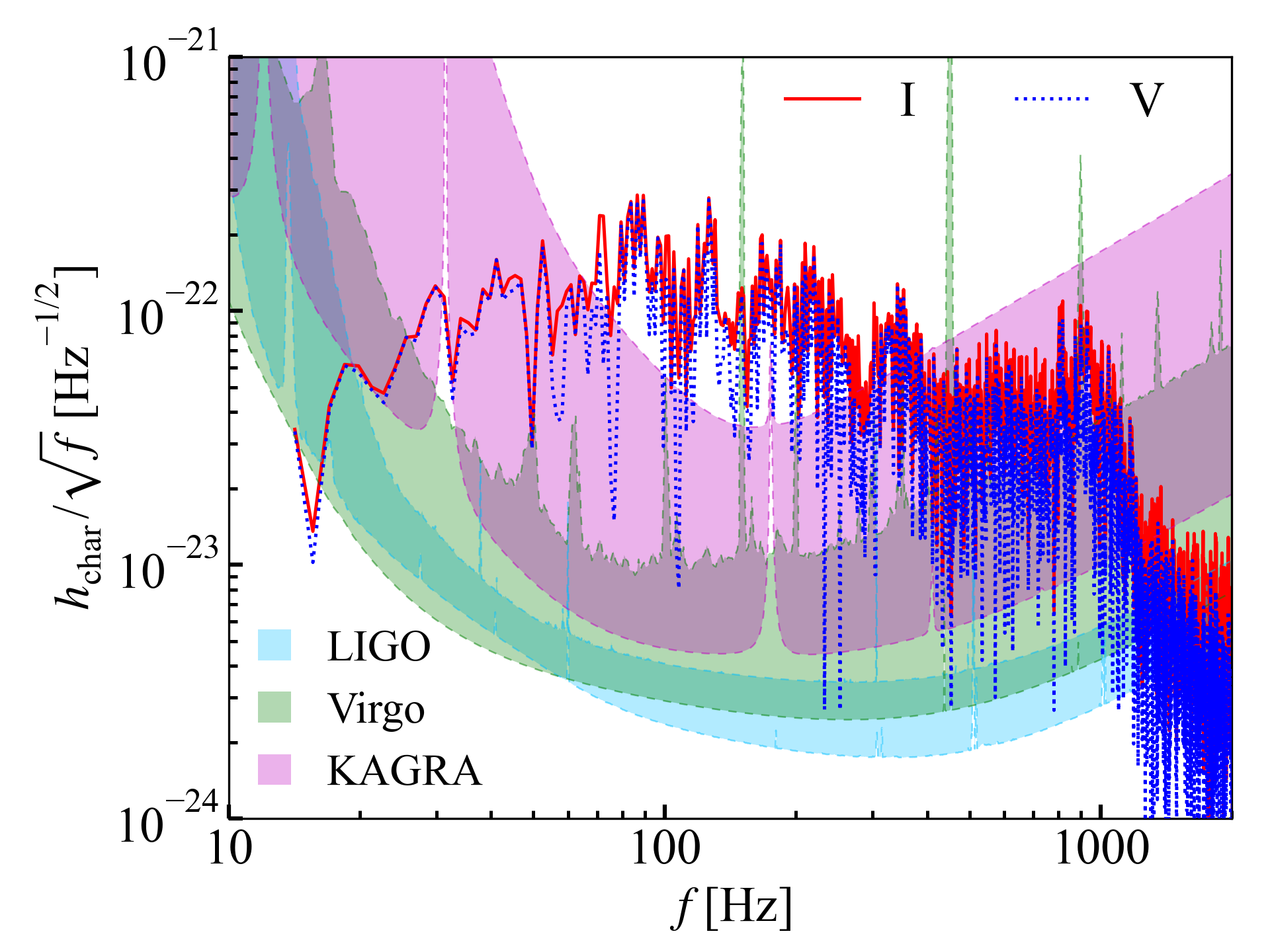}
\caption{
Amplitude spectral densities of the I-mode (solid red) and V-mode (dotted blue) GW strains for a polar observer at 10~kpc. We also show the shaded sensitivity bands of current detectors: LIGO (light blue), Virgo (green), and KAGRA (magenta), bounded by the achieved O4 and target O5 sensitivities (dashed). 
\label{fig:detectV}}
\end{figure}

In Fig.~\ref{fig:detectV} we compare the spectra of the characteristic GW strains seen from the pole at a distance of 10~kpc for the $I$ and $V$ modes 
(see Appendix~\ref{Sec:NumericalSetup} for details)
with the achieved O4 and target O5 sensitivity curves of LIGO~\citep{LIGO:T2000012,LIGO:T2500310}, Virgo~\citep{zenodoGWTC4,LIGO:T2000012}, and KAGRA~\citep{LIGO:T2000012,KAGRA:T1707038}. 
Following \citet{Shibagaki21}, we calculated the sensitivity curves of the GW detectors for the $V$ mode by assuming that two co-located detectors measure the GW plus and cross modes, respectively, and that the detector noises are Gaussian. 
Figure~\ref{fig:detectV} shows that the $I$- and $V$-mode characteristic GW strains peak at $\sim$90~Hz. 
In this idealized situation, the signal-to-noise ratios of the GW peaks for the upcoming O5 run are expected to fall between the achieved O4 and target O5 levels, reaching $\sim$50--70 for the most sensitive detectors.

\begin{figure}
\centering
\includegraphics[width=0.44\textwidth]{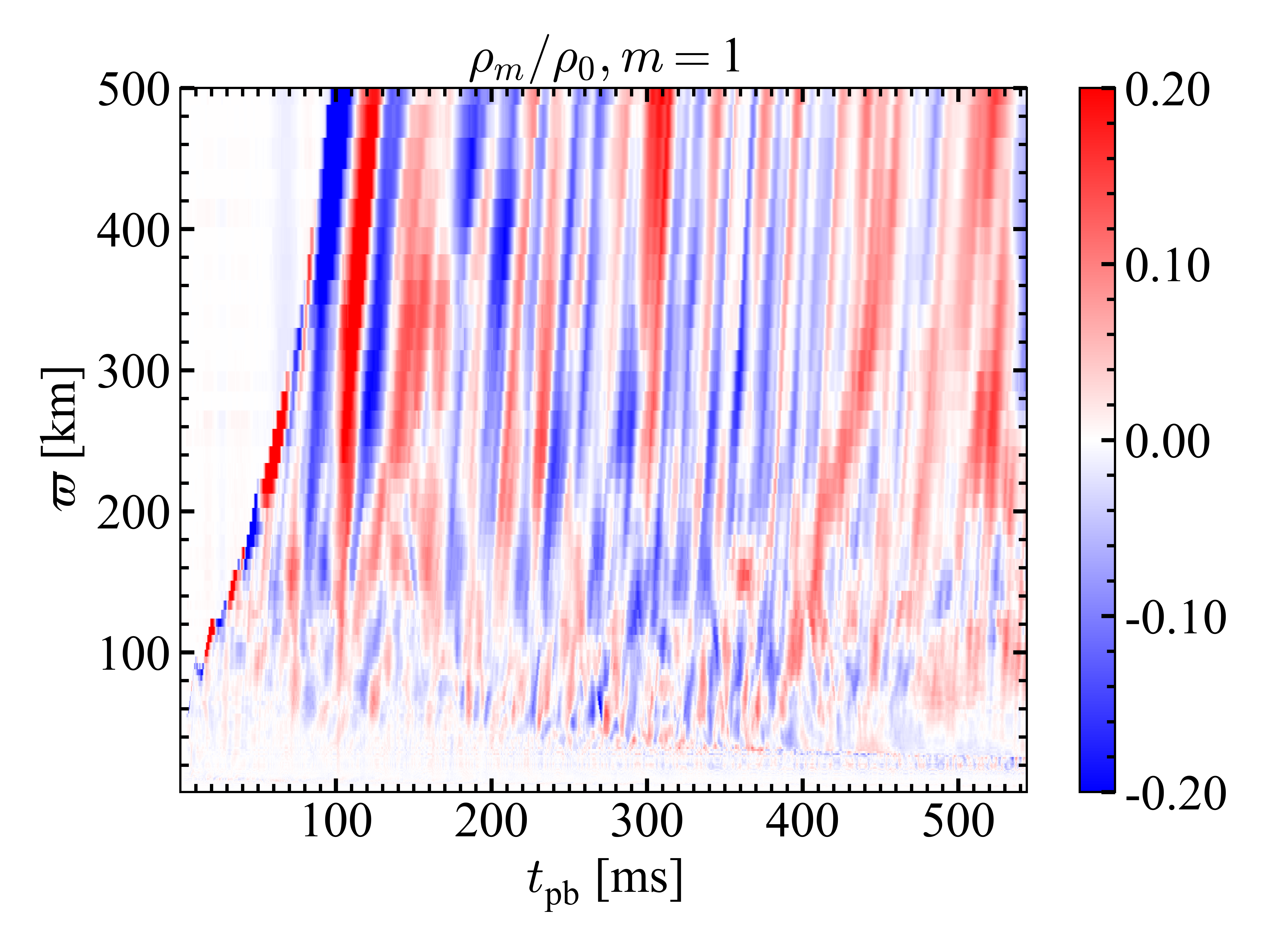}
\vspace{-2mm}
\caption{Color map of the normalized $m=1$ mode amplitude for the density deviation $\rho_m / \rho_0$ as a function of time and cylindrical radius. \label{fig:delrho}}
\end{figure}

To identify the origin of the coherent $V$-mode GW power spectrum at $t_{\rm{pb}}<230$\,ms shown in Fig.~\ref{fig:specV}, we performed a mode analysis of the density on the equatorial plane,
$\rho_{m}\left(t, \varpi \right)=\int _0 ^{2\pi} \rho(t, \varpi, \phi, z=0) \cos{(m \phi)} d\phi.$
Figure~\ref{fig:delrho} shows the $m=1$ mode amplitude relative to the $m=0$ mode amplitude ($\rho_1 / \rho_0$) in the spacetime diagram.
The largest relative amplitude is obtained at $t_{\rm{pb}}\sim 100$\,ms.
This is also evident in the density deviation of Fig.~\ref{fig:3Dent} (red and blue disk).
The regular appearance of alternating red and blue bands in Fig.~\ref{fig:delrho} indicates that a density pattern rotates simultaneously with a given mode modulation.
The finite slope of the red and blue bands refers to a nonvanishing radial velocity of the density pattern. The positive slope of the numerous bands, as shown in Fig.~\ref{fig:delrho}, indicates that the associated density patterns propagate outward.
This is a characteristic feature of the low-$T/|W|$ instability, which generates a quasi-periodic GW \citep[e.g.,][]{Shibagaki21}.
We note that this is not due to the spiral SASI because our model explodes almost without forming a standing shock.
As shown in the previous studies \citep[e.g.,][]{Takiwaki21}, the GW excited by the low-$T/|W|$ instability appears at approximately twice the local angular velocity of the fluid. In our model, the $\sim$90~Hz GW signal corresponds to a region at a cylindrical radius of $\sim$50~km, where the angular velocity is $\sim$45~Hz
(see Appendix~\ref{Sec:EmissionRegion} for a more detailed analysis).

The white region in the upper left corner of Fig.~\ref{fig:delrho} corresponds to the outside of the shock surface, so the boundary between this white region and the red or blue region corresponds to the shock radius.
After $t_{\rm{pb}}\sim100$\,ms, the speed of the shock expansion and the propagation speed of the density pattern are comparable, such that the density pattern cannot reach the shock radius. 
This indicates that the advective-acoustic cycle, a mechanism that drives the SASI \citep{Foglizzo00,Scheck08}, is not involved in the generation of these density patterns after $t_{\rm{pb}}\sim100$\,ms.

\section{Conclusions}
\label{Sec:Conclusion}
We investigated circular polarization of GWs from 3D GRMHD simulations of a magnetorotational supernova model \citep{Shibagaki24}. 
Strong circular polarization emerges along the initial rotation axis during the early post-bounce phase (up to $t_{\rm pb}\sim 230\, \mathrm{ms}$). 
This feature is associated with rotationally induced nonaxisymmetric deformations, in particular, the development of the low-$T/|W|$ instability and spiral arm structures \citep[e.g.,][]{Ott05,Shibagaki21}.  

The characteristic GW spectra 
are found to peak around $90\,\mathrm{Hz}$, which can be interpreted as twice the local angular velocity of the fluid in the region where $\Omega\sim 45\,\mathrm{Hz}$ at a cylindrical radius of $\sim$50\,km. 
This interpretation is consistent with the theoretical expectation that the low-$T/|W|$ instability excites GWs near twice the rotation frequency \citep[e.g.,][]{Takiwaki21}. 
The frequency range of $\sim$90~Hz falls within the sensitivity bands of current ground-based detectors such as Advanced LIGO, Advanced Virgo, and KAGRA, suggesting a potential for detectability under idealized noise assumptions.

These results highlight the potential of GW polarization measurements as a diagnostic tool for probing the CCSN explosion mechanism and the dynamical PNS evolution. 
While previous studies mainly associated strong circular polarization with rapidly rotating nonmagnetized collapses \citep{hayama16,Shibagaki21}, our findings have demonstrated that models that launch MHD jets can also exhibit distinct polarization signatures. 
Importantly, the circularly polarized GWs identified here originate predominantly from the low-$T/|W|$ instabilities and nonaxisymmetric motions around the PNS surface and are not directly generated by the MHD jets themselves. This distinction suggests that future detector networks, with enhanced polarization analysis capabilities, can exploit these features to distinguish between different explosion scenarios.

Several limitations of this work should be mentioned. 
Our conclusions are based on a single progenitor model 
of a zero-age main-sequence mass of $20 \, M_\odot$ with rapid rotation and a strong magnetic field. A systematic exploration of the progenitor diversity, including variations in rotation rates and magnetic field strengths, will be necessary to confirm the robustness of these findings. Moreover, the role of turbulence and nonlinear instabilities requires further study, particularly with respect to numerical resolution and diffusion effects \citep{Moesta14,Bugli21}. Finally, a more realistic evaluation of detectability will require incorporating detector noise properties and advanced data-analysis strategies.  

In summary, we identified circularly polarized GW emission as a robust feature of magnetorotational explosions and provided a pathway toward using polarization as a probe of the central engine in energetic core-collapse events. 
Our findings serve as a milestone for future studies, including broader parameter surveys and predictions for third-generation detectors.
These will be crucial to fully exploit the diagnostic potential of GW polarization in multimessenger astrophysics.

\begin{acknowledgements}
Numerical computations were carried out on Cray XC50 at the Center for Computational Astrophysics, National Astronomical Observatory of Japan and on Cray XC40 at YITP in Kyoto University. This work was supported by 
JSPS KAKENHI
Grant Number 
(JP17H06364, 
JP22H01223,
JP21H01088, 
JP23H01199, 
JP23K03400,
JP23K22494,
JP24K00631,
and JP26K07093),
MEXT as “Program for Promoting researches on the Supercomputer Fugaku” (Structure and Evolution of the Universe Unraveled by Fusion of Simulation and AI; Grant Number JPMXP1020230406), and JICFuS.
The authors acknowledge support from the program Excellence Initiative--Research University of the University of Wrocław of the Ministry of Education and Science and the Scultetus Visiting Scientist Program of the Center for Advanced Systems Understanding (S.S.) and the Polish National Science Centre (NCN) under grant number 2023/49/B/ST9/03941 (S.S. and T.F.) as well as by the WCSS Wrocław Centre for Scientific Computing and Networking (T.F.).
\end{acknowledgements}

   \bibliographystyle{aa}
   \bibliography{mybib}

\begin{appendix}
\section{Numerical setup}
\label{Sec:NumericalSetup}
\subsection{Simulation code and initial condition}
We carried out our simulations with the 3D GR neutrino-radiation ideal MHD code developed by \citet{KurodaT21}. The spacetime metric is evolved using the Baumgarte–Shapiro–Shibata–Nakamura formulation \citep[see, e.g.,][]{Shibata95,Baumgarte99,Marronetti08} on a fixed Cartesian mesh. As in \citet{KurodaT21}, the grid resolution is chosen such that the finest level near the center reaches $\Delta x = 458$\,m. The computational domain covers a region of $1.5\times10^4$\,km in radius, with 10 nested refinement levels in a 2:1 ratio. Each refinement box contains $64^3$ zones.  

As the progenitor, we use the solar-metallicity $20 \, M_{\odot}$ model ``s20a28n'' from \citet{WH07}, which is commonly employed in CCSN studies. A cylindrical rotation law is imposed, setting the initial angular momentum distribution as $u^t u_{\phi} = \varpi_0^2 \, (\Omega_0 - \Omega)$, 
where $u^t$ is the temporal component of the four-velocity, $u_{\phi} \equiv \varpi^2 \Omega$ with $\varpi = \sqrt{x^2+y^2}$, and $\varpi_0$ is taken to be $10^8$\,cm.  

The initial magnetic field is specified through a purely toroidal vector potential, $A_{\phi} = \frac{B_0}{2} \, \frac{R_0^3}{r^3+R_0^3} \, r \sin\theta$, $A_r = A_\theta = 0$, 
which yields an approximately uniform vertical field inside $r < R_0$ and a dipolar field outside. In this study we set $R_0 = 10^8$\,cm.  For the present work we analyze the model with $(\Omega_0\,[\mathrm{rad\,s^{-1}}], B_0[\mathrm{G}]/\sqrt{4\pi}\,) = (2.0, 10^{12})$, denoted as R20B12 following \citet{Shibagaki24}.

\subsection{Gravitational wave analysis}
We extract the GWs with a standard quadrupole formula \citep{Shibata&Sekiguchi03,KurodaT14}. To investigate the spectral evolution of the circularly polarized component of the GWs, we evaluate the Stokes I and V parameters \citep{Seto07,hayama16},
\begin{align}
I\left(f \right) &= \frac{\tilde{h}_{R}\tilde{h}_{R}^* +  \tilde{h}_{L} \tilde{h}_{L}^*}{2} \Delta f = \frac{\tilde{h}_{+}\tilde{h}_{+}^* +  \tilde{h}_{\times} \tilde{h}_{\times}^* }{2}\Delta f, \label{eq:stokesI} \\
 V\left(f \right) &= \frac{\tilde{h}_{R}\tilde{h}_{R}^* - \tilde{h}_{L} \tilde{h}_{L}^*}{2} \Delta f = i \frac{ \tilde{h}_{+} \tilde{h}_{\times}^* -\tilde{h}_{\times} \tilde{h}_{+}^* }{2}\Delta f, \label{eq:stokesV}
\end{align}
where $\tilde{h}$ is the Fourier amplitude of the GW strain, $h$, and $\Delta f$ is the smallest frequency width of the Fourier transform, i.e., the inverse of the integration time for the Fourier transform.
The $h_R$ and $h_L$ are the right-handed and left-handed polarization modes, which are defined by $h_R = (h_+-ih_{\times})/\sqrt{2}$ and $h_L = (h_++ih_{\times})/\sqrt{2}$.
The Stokes I parameter indicates the total power spectrum, while the Stokes V parameter represents the power spectrum of the difference between its clockwise circularly polarized component and counterclockwise one.
For the discussion of the detectability of these modes, it is convenient to define the characteristic GW strain for the $I$ and $V$ modes,
\begin{align}
  h_{\rm{char},\it{I}}^2 &= \frac{8f^{2} I}{\Delta f}, \label{eq:hchar_I}\\
  h_{\rm{char},\it{V}}^2 &= \frac{8f^{2} |V|}{\Delta f}. \label{eq:hchar_V}
\end{align}
These quantities are defined so that $h_{\rm{char},\it{I}}$ becomes identical to the well-known characteristic GW strain \citep{Shibagaki21}.

To perform time-frequency analysis of these modes, we perform a short-time Fourier transform of the GW strain with the Hann window:
\begin{align}
H\left(t, f \right) &= \int^{t+\Delta t/2}_{t-\Delta t/2} h\left(\tau\right) W(\tau-t) e^{2 \pi i f \tau}d\tau, \label{eq:STFT}\\
 W\left(x \right) &=
\left\{ 
\begin{array}{ll}
 \frac{1}{2}\left(1+\cos{\left(\frac{2 \pi x}{\Delta t}\right)}\right) & \left(|x| < \Delta t / 2 \right)\\
 0 & \left(|x| \geq \Delta t / 2 \right)
 \end{array}
 \right.. \label{eq:window}
\end{align}
Replacing $\tilde{h}$ in Eqs.\,(\ref{eq:stokesI}) and (\ref{eq:stokesV}) with $H$ in Eq.\,(\ref{eq:STFT}), we can obtain time-frequency spectrograms of the $I$ mode, $I\left(t,f\right)$, and the $V$ mode, $V\left(t,f\right)$.
In this study, we use 20\,ms of $\Delta t$.

\section{Emission Region}\label{Sec:EmissionRegion}
\begin{figure}[hbt]
\centering
\includegraphics[width=0.475\textwidth]{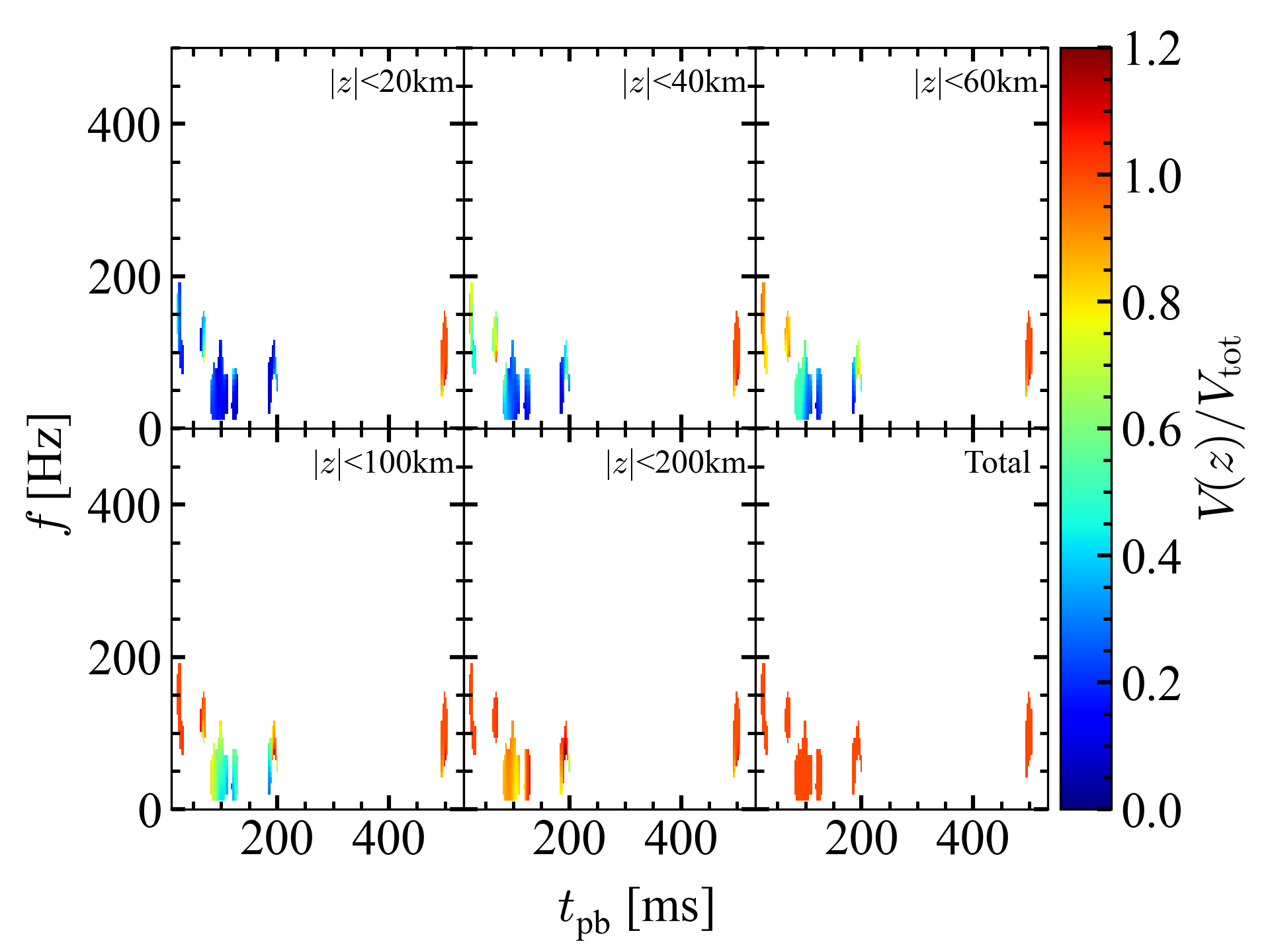}
\caption{Relative contributions from each rectangular box defined by $|z|<z_0$ to the total $V$-mode GW spectrogram (seen from the pole).
We color only the $t-f$ domains with $|V_{\rm tot}|>10^{-46}$.
The value of $z_0$ is shown in the upper right corner of each panel.
As a reference, the bottom right panel shows the total $V$-mode GW spectrogram.  \label{fig:Vzlayer}}
\end{figure}

For further analysis of the GW emission regions, we compute $V$-mode spectrograms observed from the pole using quadrupole moments of limited spatial domains ($|z|<z_0$). 
Figure~\ref{fig:Vzlayer} shows the $V$-mode spectrograms normalized by the total $V$-mode amplitude.
$z_0$ is denoted on  the upper right corner.
To extract important parts of the spectrograms, we only plot regions where the total $V$-mode spectrogram is larger than $10^{-46}$.
Looking at $t_{\rm{pb}}<230$\,ms, $\sim$60\% of the $V$-mode amplitude is generated in the region of $|z|<100$\,km and $\sim$80\% in the region of $|z|<200$\,km.
This is consistent with our interpretation that the GWs are generated by not only the deformed PNS but also extended spiral arms driven by the low-$T/|W|$ instability.

\end{appendix}

\end{document}